%% file: main.tex
\PassOptionsToPackage{table}{xcolor}
\documentclass[letterpaper,twocolumn,10pt]{article}
\usepackage{usenix}

\usepackage{amsmath,amssymb,amsfonts}
\usepackage{amsthm}
\usepackage{bm}
\usepackage{graphicx}
\usepackage[table]{xcolor}
\usepackage{booktabs}
\usepackage{tabularx}
\usepackage{threeparttable}
\usepackage{multirow}
\usepackage{makecell}
\usepackage{enumitem}
\usepackage{stmaryrd}
\usepackage{framed}
\usepackage{caption}
\usepackage{subfigure} 
\usepackage{pifont}
\usepackage{dashrule}
\usepackage{tikz,pgfplots}
\usetikzlibrary{matrix,fit,calc,patterns,backgrounds,arrows,shapes,chains,decorations,intersections,positioning}
\usepackage[title]{appendix}

\title{\Large \bf UniMark: Artificial Intelligence Generated Content Identification Toolkit}
\author{
\textbf{Meilin Li}\textsuperscript{1,2}, \textbf{Ji He}\textsuperscript{1,3}, \textbf{Yi Yu}\textsuperscript{1}, \textbf{Jia Xu}\textsuperscript{1}, \textbf{Shanzhe Lei}\textsuperscript{1}, \textbf{Yan Teng}\textsuperscript{1}, \textbf{Yingchun Wang}\textsuperscript{1}, \textbf{Xuhong Wang}\textsuperscript{1}\thanks{Corresponding author.}\\
\textsuperscript{1}\textrm{Shanghai AI Laboratory}, \textsuperscript{2}\textrm{Shandong University}\\
\textsuperscript{3}\textrm{Shanghai Information Security Testing Evaluation and Certification Center}\\
\texttt{lml@mail.sdu.edu.cn}, \texttt{heji@shtec.org.cn}\\
\texttt{\{yuyi, xujia, leishanzhe, tengyan, wangyingchun, wangxuhong\}@pjlab.org.cn}
}
\date{}

\begin{document}
\maketitle

\begin{abstract}
The rapid proliferation of Artificial Intelligence Generated Content has precipitated a crisis of trust and urgent regulatory demands. However, existing identification tools suffer from fragmentation and a lack of support for visible compliance marking. To address these gaps, we introduce the \textbf{UniMark}, an open-source, unified framework for multimodal content governance. Our system features a modular unified engine that abstracts complexities across text, image, audio, and video modalities. Crucially, we propose a novel dual-operation strategy, natively supporting both \emph{Hidden Watermarking} for copyright protection and \emph{Visible Marking} for regulatory compliance. Furthermore, we establish a standardized evaluation framework with three specialized benchmarks (Image/Video/Audio-Bench) to ensure rigorous performance assessment. This toolkit bridges the gap between advanced algorithms and engineering implementation, fostering a more transparent and secure digital ecosystem. The code is available at 
\url{https://github.com/AI45Lab/UniMark}. 
\end{abstract}

\input{1.Introduction.tex}
\input{3.Conclusion.tex}


\appendix
\bibliographystyle{plain}
\bibliography{main}

\end{document}

%% file: 1.Introduction.tex
\section{Introduction}

  In recent years, Large Language Models (LLMs) and generative diffusion models have made remarkable progress. From text-domain models such as GPT-5 \cite{gpt4o} and Qwen \cite{qwen2}, to video generation models like Sora 2 \cite{sora}, Wan 2.1 \cite{wan}, and Nano Banana \cite{nano}, Artificial Intelligence Generated Content (AIGC) is reshaping the production of digital content at an unprecedented pace. However, the widespread adoption of these technologies has also precipitated a profound crisis of trust. The abuse of Deepfake technology, the viral spread of fake news, and increasingly severe copyright infringement issues are eroding the foundation of public trust in digital information. Facing this challenge, establishing a reliable mechanism for Content Identification and Tracing is no longer optional but an urgent necessity for the digital ecosystem. Governments worldwide have also attached great importance to this issue; for instance, China's national standard \emph{Information Security Technology -- Method for Identifying Content Generated by Artificial Intelligence} \cite{gb} and the European Union's AI Act \cite{eu_ai} have both set forth clear specifications and requirements for the identification of AIGC.

  Existing AIGC identification solutions fall primarily into two categories: \emph{In-processing} (generation process embedding) and \emph{Post-processing} (generation after-treatment). In-processing methods embed watermarks directly during the model inference stage. Representative works include SynthID \cite{synthid} and GumbelSoft \cite{gumbelsoft} in the text domain; Stable Signature \cite{stable_signature}, Tree-Ring \cite{tree_rings}, and PRC \cite{prc} in the image domain; and Groot \cite{groot} in the audio domain. Although these methods theoretically offer high imperceptibility, they face significant limitations in practical application: they typically require access to internal model parameters (such as logits), making them inapplicable to black-box APIs like GPT-5 or Midjourney, and they cannot handle pre-existing generated data. Consequently, to build a universal and model-agnostic solution, Post-processing schemes have become a more practical choice. Related research, such as PostMark \cite{postmark} for text, ZoDiac \cite{zodiac} for images, VideoSeal \cite{videoseal} for video, and AudioSeal \cite{audioseal} for audio, has demonstrated immense potential. However, the current ecosystem of identification tools suffers from severe fragmentation—State-of-the-Art (SOTA) algorithms are scattered across different codebases with inconsistent interface standards, making integration difficult. More importantly, existing tools mostly focus solely on implicit technical watermarks, neglecting ``Visible Marking''—a critical dimension for regulatory compliance—thus failing to meet the requirements for user right-to-know mandated by regulations such as the EU AI Act.

To address these gaps, we propose the \textbf{UniMark}—an open-source, modular, and production-grade multimodal content identification framework.

First, the core of this framework is a unified engine architecture. We have highly abstracted the differences among text, image, audio, and video modalities. Developers need not learn complex underlying libraries for different modalities; they can complete all operations through unified \texttt{embed} and \texttt{extract} APIs, significantly lowering the barrier to entry for multimodal application development.

Second, addressing the dual needs of copyright protection and regulatory compliance, we propose a unique dual-operation strategy. This strategy natively supports both implicit \emph{Watermarks} for technical tracing and \emph{Visible Marks} for compliance requirements—such as video overlays or audio voice prompts—truly achieving the unification of technical protection and legal compliance.

Finally, considering the rapid iteration of AIGC algorithms, we designed the architecture as a high-extensibility algorithm platform. By defining standardized interface protocols, researchers can integrate the latest SOTA algorithms, such as PostMark \cite{postmark} and CredID \cite{credid}, into the framework at zero cost, enabling it to evolve continuously with technological developments.

To resolve the issue of inconsistent evaluation standards, this toolkit incorporates three standardized benchmarks: Image-Bench (based on the W-Bench dataset), Video-Bench (based on the VideoMarkBench dataset), and Audio-Bench (based on the AudioMark dataset). The main contributions of this paper are as follows: 
\begin{itemize}
    \item We propose a unified multimodal framework that abstracts the complexity of text, image, audio, and video watermarking into a single engine, significantly lowering the barrier to entry for developers.
    \item We introduce a novel dual-operation strategy that natively supports both hidden watermarking for technical tracing and visible marking for regulatory compliance, addressing the dual needs of copyright protection and user transparency.
    \item We establish a standardized evaluation framework comprising Image-Bench, Video-Bench, and Audio-Bench, providing the community with a rigorous and unified benchmark for assessing algorithm performance across diverse modalities.
\end{itemize}

\section{System Design}\label{system_design}

\subsection{Unified Engine Architecture}
The core of the \textbf{UniMark} is the \texttt{UnifiedWatermarkEngine}, designed to abstract the complexities of multimodal processing into a cohesive framework. We employ the Facade Pattern through the \texttt{WatermarkTool} class, providing developers with a simplified, high-level interface that masks the intricate details of underlying model initialization and tensor manipulations.

\begin{figure}[h]
    \centering
    \includegraphics[width=0.95\linewidth]{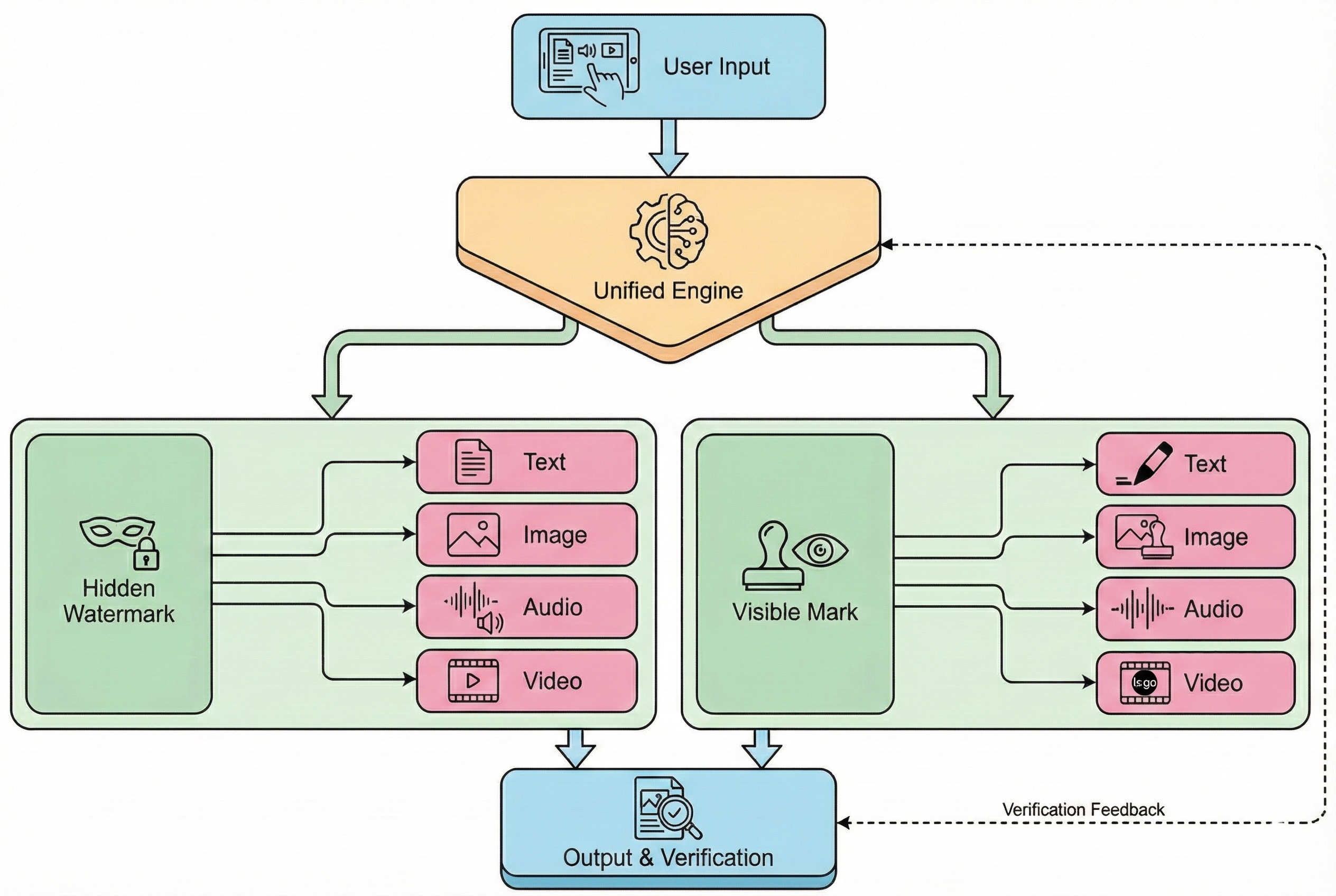}
    \caption{\textbf{System Architecture of UniMark.} The Unified Engine acts as a central hub, routing user inputs through a dual-operation pathway (Hidden Watermark vs. Visible Mark) across four modalities, ensuring both copyright protection and regulatory compliance.}
    \label{fig:architecture}
\end{figure}

A key innovation in our architecture is the Dual-Operation Strategy, as illustrated in Figure \ref{fig:architecture}. Unlike traditional systems that focus solely on invisible steganography, our engine natively supports two distinct operation modes via unified \texttt{embed} and \texttt{extract} APIs:
\begin{itemize}
    \item \textbf{Hidden Watermarking} (\texttt{operation='watermark'}): Embeds imperceptible signals for copyright protection and technical tracing, utilizing deep learning-based encoders.
    \item \textbf{Visible Marking} (\texttt{operation='visible\_mark'}): Applies explicit compliance markers (e.g., visual overlays or audio prompts) to meet regulatory transparency requirements, such as those mandated by the EU AI Act \cite{eu_ai}.
\end{itemize}

To ensure efficiency and scalability, the system implements a Lazy Loading mechanism. Heavy resources, such as the 1.3B parameter Wan 2.1 video generation model or the AudioSeal encoder, are instantiated only upon the first request for their specific modality. Furthermore, the entire system is Config-Driven, managed by a centralized YAML configuration (\texttt{config/default\_config.yaml}), allowing researchers to dynamically switch algorithms or adjust hyperparameters without modifying the source code.

\subsection{Multimodal Algorithm Integration}
The toolkit integrates State-of-the-Art (SOTA) algorithms across four modalities through a standardized adapter interface. In this section, we primarily detail the integration of the visual modality, followed by a brief overview of the audio and text modalities.

For both image and video modalities, we unify the backend using VideoSeal \cite{videoseal}. By treating images as single-frame videos, we leverage a single pre-trained encoder to embed watermarks that are robust against both spatial editing (e.g., cropping, resizing) and temporal manipulation (e.g., frame dropping). This approach significantly reduces code redundancy and maintenance overhead while ensuring consistent performance across visual content. Furthermore, we construct an automated pipeline that seamlessly links generative models with the watermarking process; for instance, video content generated by Wan 2.1 \cite{wan} is automatically passed to the VideoSeal encoder for immediate protection.

Regarding other modalities, we integrate AudioSeal \cite{audioseal} for audio watermarking, which enables precise identification of AI-generated segments within longer streams through localized detection. For text, we employ PostMark \cite{postmark} to address the prevalence of closed-source LLMs. Unlike generation-time methods that require access to model logits, PostMark operates as a post-processing step, ensuring our toolkit remains model-agnostic and compatible with black-box APIs such as GPT-4.

\subsection{Standardized Evaluation Framework}
To address the fragmentation in evaluation standards, we provide a comprehensive benchmarking suite located in the \texttt{benchmarks/} directory. Its primary objective is to evaluate the performance of watermark algorithms across different modalities by providing standardized test datasets, diverse attack methods, and unified evaluation metrics, thereby assisting users in selecting the most suitable algorithm for their specific application scenarios.

\begin{figure}[h]
    \centering
    \includegraphics[width=0.95\linewidth]{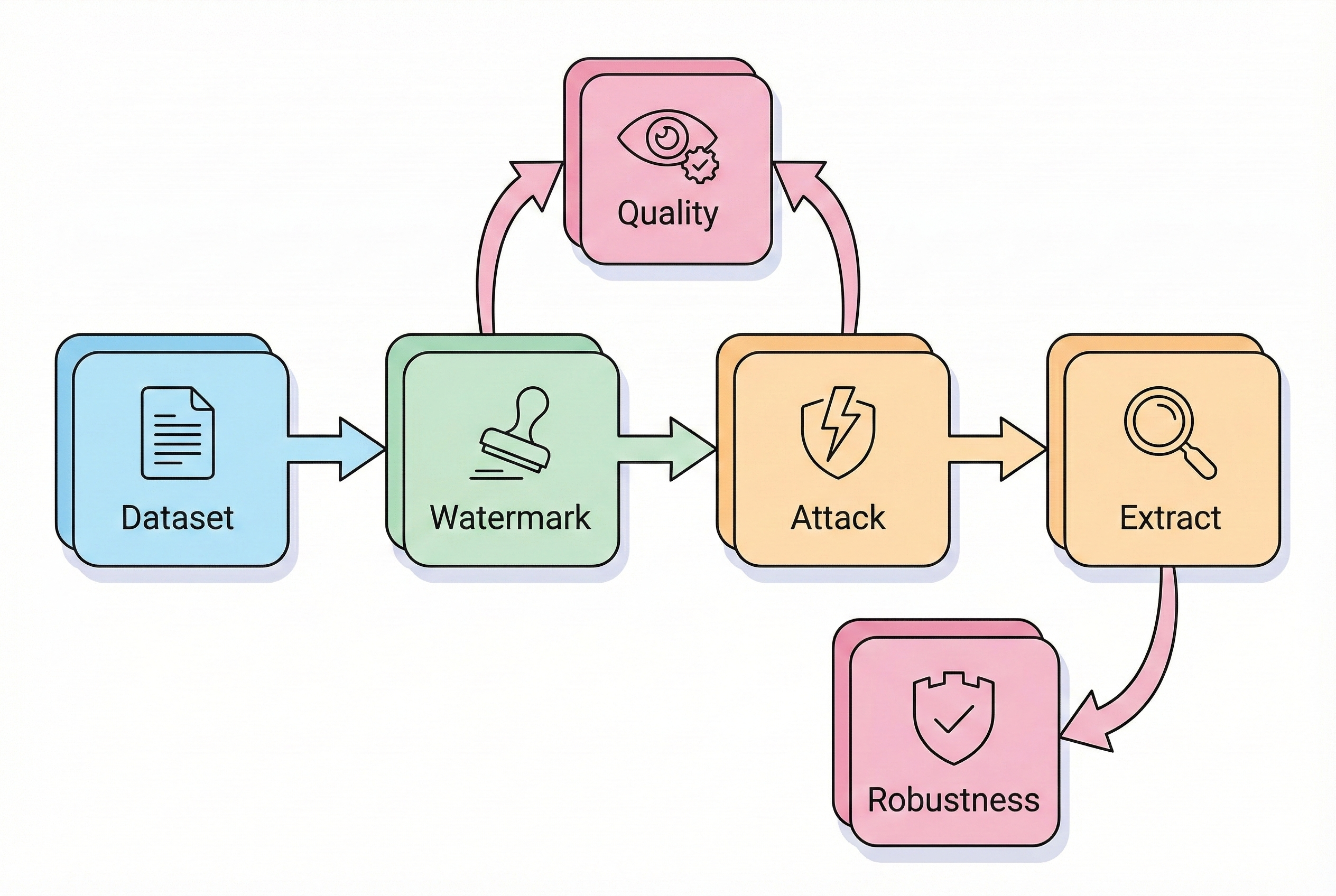}
    \caption{\textbf{Evaluation Pipeline of UniMark.} The process follows a rigorous "Attack-Detect-Score" loop: starting with a standardized Dataset, content undergoes Watermark embedding. Quality is evaluated by comparing watermarked content with the original. Subsequently, Attack simulations (e.g., noise, compression) are applied, followed by watermark Extraction. Finally, Robustness is assessed by comparing the extracted message against the original.}
    \label{fig:eval_pipeline}
\end{figure}

We offer three specialized benchmarks covering the major modalities. \textbf{Image-Bench}, built upon the W-Bench dataset, evaluates robustness against 25 types of distortions, including common operations like JPEG compression and Gaussian blurring. \textbf{Video-Bench}, based on the VideoMarkBench dataset, focuses on temporal consistency and resistance to frame-level attacks such as frame dropping and averaging. \textbf{Audio-Bench} utilizes the AudioMark dataset to assess robustness against audio-specific perturbations like noise injection, time-stretching, and MP3 compression.

The framework evaluates algorithms across two critical dimensions: Quality and Robustness. Quality metrics measure the imperceptibility of the watermark, utilizing PSNR, SSIM, and LPIPS for visual content, and SNR for audio, to ensure that the watermarking process does not degrade the user experience. Robustness metrics quantify the detectability of the watermark under various attacks, primarily using True Positive Rate (TPR) and Bit Accuracy (BA). The \texttt{BenchmarkRunner} automates the calculation of these metrics, generating standardized reports and radar charts to facilitate fair and intuitive comparisons between different algorithms.

%% file: 3.Conclusion.tex
\section{Conclusion}

In this work, we presented the \textbf{UniMark}, a comprehensive solution designed to address the fragmentation and lack of standardization in the current landscape of AI-generated content governance. By introducing a unified engine, we have significantly lowered the barrier to entry for integrating multimodal watermarking technologies, enabling developers to access state-of-the-art algorithms through a single, consistent interface. Our novel dual-operation strategy successfully bridges the gap between technical tracing and regulatory compliance, offering a unified pathway for both hidden copyright protection and visible user transparency. Furthermore, the inclusion of a standardized evaluation framework provides the community with a rigorous set of benchmarks, fostering fair comparison and accelerating the development of more robust identification methods.

Looking ahead, we plan to expand the toolkit's capabilities to further enhance the trustworthiness of the digital ecosystem. A key direction is the implementation of \textbf{Implicit Metadata Marking}, which will allow for the embedding of structured metadata—such as model versions and generation timestamps—directly into the content, thereby enriching the granularity of provenance tracking. We also aim to continuously integrate emerging SOTA algorithms to maintain the toolkit's relevance in this rapidly evolving field. We invite the open-source community to collaborate with us in building a more transparent and secure future for AI-generated content.